\documentstyle[pre,aps]{revtex}
\draft
\begin{document}
\title{Band Formation during Gaseous Diffusion in Aerogels}
\author{M.\ A.\ Einarsrud}
\address{Institutt for uorganisk kjemi, Norges teknisk-naturvitenskapelige 
universitet, N--7034 Trondheim, Norway}
\author{Frank A.\ Maa{\o} and Alex Hansen}
\address{Institutt for fysikk, Norges teknisk-naturvitenskapelige 
universitet, N--7034 Trondheim, Norway}
\author{M.\ Kirkedelen and J.\ Samseth}
\address{Institutt for energiteknikk, Postboks 40, N--2044 Kjeller, Norway}
\date{\today}
\maketitle
\begin{abstract}
We study experimentally how gaseous HCl and NH$_3$ diffuse from opposite 
sides of and react in silica aerogel rods with porosity of 92 \% and average 
pore size of about 50 nm.  The reaction  leads to solid NH$_4$Cl, which is 
deposited in thin sheet-like structures.
We present a numerical study of the phenomenon.  Due to the 
difference in boundary conditions between this system and those usually 
studied, we find the sheet-like structures in the aerogel to differ 
significantly from older studies.  The influence of random nucleation
centers and inhomogeneities in the aerogel is studied numerically.
\end{abstract}
\section{Introduction}
\label{sect1}

A classic high-school chemistry experiment consists of placing a cotton plug 
drenched in ammonia at one end of a long glass tube simultaneously with 
another one drenched in hydrochloric acid at the other end of the 
tube\cite{z95}.  Then one waits, and after some time, a white ring forms on
the tube wall. The white ring consists of ammonium chloride, resulting from 
the gases reacting on contact.  From 
measuring the position of the ring relative to the two ends of the tube, the 
ratio between the average velocities of the two gases is found. This 
ratio is then compared to Graham's law which states that it is equal to the 
square root of the inverse of the molar masses of the two gases. 

What happens if we repeat this experiment with a porous medium substituting 
for the air-filled tube? We have performed such experiments, using a silica 
aerogel as the porous medium.  A large number of closely spaced 
paper-thin sheets form in the aerogel, spanning it in the radial 
direction.  In Figure \ref{fig1}, we show a photograph of the precipitate 
that was formed by exposing the aerogel rod during 5.5 days to the reacting 
gases.   

Periodic sheet-like structures are known to develop in diffusion-reaction 
systems.  They were first described one hundred and one years ago by Liesegang
\cite{l96}, who observed the reaction when silver nitrate solution diffuses 
into a gel containing silver dichromate.  About a year later, Ostwald 
\cite{o97} suggested that the Liesegang rings are due the presence of a
nucleation threshold.  The reaction product nucleates only when a threshold
concentration is reached.  This nucleation depletes the concentration of mobile
reaction product in the zone where the concentration is above threshold and
its neighborhood, thus stopping the nucleation process here.  Meanwhile, the
reaction front moves on, building up the concentration of mobile reaction 
product elsewhere.  This leads to the formation of the Liesegang rings.
Another theory, put forward by Prager \cite{p56}, is based on the 
existence of a reaction threshold between the two diffusing species.
An intermediate mobile reaction product is then no longer necessary in
order to produce the Liesegang structure, as was the case in the Ostwald
theory.  

In the subsequent years, hundreds of papers have appeared discussing various
aspects of the Liesegang phenomena, including a host of alternative 
explanations. For reviews see e.g.\ \cite{s54,kr81,cld94,kbz95}.  
We note that in standard
Liesegang experiments, one of the reactants is already present in the gel 
at the beginning of the experiment. However, in our experiment the two 
reactants simultanously diffuse into the reaction zone. 

Usually, the reactants diffuse in an aqueous
gel.  However, there exist experiments that have demonstrated the Lieseang
phenomenon in gaseous systems, notably that of Spotz and Hirschfelder 
\cite{sh51} who obtained rings in a tube containing HCl and NH$_3$, i.e.,
the high-school setup described above. As no porous medium was deployed in 
this study, the reaction product were not kept fixed and the structure of the 
rings could not be studied.

In the next section we describe the fabrication 
process of the aerogels used in the experiments
and present the results of these. Besides the visual observation of the 
Liesegang rings, we measure effective diffusion constants and use 
IR-spectroscopy to more accurately determine
the concentration profile of 
the precipitate. This latter is important since the pore size is
about a tenth of the wave length of 
visible ligth, making it impossible to assess the true shape
of the Liesegang sheets.
In section \ref{sect3} we present a numerical study of a reaction 
diffusion system using
a slightly modified Oswald theory resembling the one studied by 
Dee \cite{d86}, and with boundary conditions similar to those used in
the experiment.
The resulting structure of the precipitate is qualitatively very similar
to those observed in the experiments.
We furthermore study the influence of random nucleation centers and 
inhomogeneities 
in the porous medium.  Our conclusions are presented in section \ref{sect4}.

\section{Experiment}
\label{sect2}
\subsection{Fabrication of aerogel}
\label{sect2a}

Silica aerogels are ideal systems to study gaseous diffusion-reaction 
processes.  They are highly porous, with a porosity up to 99.8\%.  
SANS studies \cite{vwpc89} reveal a fractal pore structure 
over a range from 0.5 nm to 50 nm. This is of the order of the molecular 
mean free path of a gas at standard pressure and temperature. The fractal 
dimension of the aerogels is in the range of 2.2 to 2.4.  At larger scales 
the gel is uniform. A micrograph presented in \cite{cmr96} shows a gel 
structure resembling a random fibrious network. The silica aerogels used in 
the present study were made from tetramethoxysilane (TMOS), H$_2$O, 
methanol, HCl and NH$_4$OH  in the total molar ratio 
$1:4.98:12.6:10^{-3}:3.8 10^{-3}$  (for some gels 
$1:4.98:12.6:10^{-3}:1.9 10^{-3}$) 
following a two-step acid-base catalyzed route described by Brinker 
{\it et al.\/} \cite{bksaka84}. To prepare hydrophobic aerogels which were 
suggested not to chemically interfere with the diffusing gases, the gels 
were treated with hexamethyldisilazane (HMDZ) in a heptane solution prior to 
drying. During this treatment a methylation of the gel surface occurs by 
substituting the OH groups present on the gel surface. The methylation of the 
gel surface makes it possible to obtain monolithic aerogels at ambient 
pressure (supercritical conditions are normally necessary) due to a 
``springback" of the gels during drying \cite{dsb94}. However, from 
spectroscopic investigations, we note that a small number of remaining OH 
groups on the pore walls still might act as hydrophilic sites in the system. 
The bulk density of the aerogels used in this work range from 0.158 g/cm$^3$ 
to 0.196 g/cm$^3$, corresponding to porosities of 
92.8\% and 91.1\%.  The aerogels were cast into rods of diameter 8.8 mm, 
giving an aerogel diameter of about 8 mm. During the diffusion experiments 
the aerogel rods were covered with a Teflon coating, forcing the diffusing 
gases to enter the aerogel only through the ends.

\subsection{Measurments of the effective diffusion coefficients}
\label{sect2b}

In order to determine the diffusion coefficients of the two gases HCl and 
NH$_3$ in the aerogel, we connected the aerogel rod in series with long 
glass tube.  Cotton plugs drenched in ammonia solution and hydrochloric acid, 
respectively, were then simultaneously placed at the open ends of the gel 
rod and the glass tube. By measuring where the NH$_4$Cl first appears in 
the glass tube, the diffusion coefficients in the aerogel may be found 
when the speeds at which the gases move in an air filled glass tube are 
known. These velocities we determine by placing cotton plugs drenched in 
ammonia and hydrochloric acid simultaneously in the ends of a long, empty 
tube --- i.e.\ the high school setup described in the introduction. In the 
glass tube of length 1.5 m we determine the ratio between the distances 
from the ends of the tube to the point at which the NH$_4$Cl precipitate 
first occurs. We found it to be 2.25. 

It should be noted that this result is very 
different from that predicted by use of Graham's law, which is 1.47.  The 
reason for this is that chemisorbtion on the surface of the glass tube 
is important. The precipitate appears after 30 minutes.  Thus, combining 
these two pieces of information, we determine the effective velocities
of the two species in the tube to be $v_{HCl} = 920$ mm/h and 
$v_{NH_3}=2070$ mm/h. Our determination of the diffusion coefficients in the 
aerogels using this method are summarized in Table \ref{tab1}.  
We would have expected samples 1 and 2 to produce equal diffusion constants 
as the same aerogel is used in these two measurements.  The same comment 
apply to samples 4 and 5.  However, they differ by approximately 15\%.  
It is the shorter sample that leads to the smaller diffusion constant for 
pair 1 and 2, while it is the longer sample in pair 4 and 5 that 
give the smaller diffusion coefficient. As a result of this lack of 
systematic trends in the data, we attribute the differences to the accuracy 
of the method we have used rather than a real effect. We note that the 
ratio $D_{NH_3}/D_{HCl}=1.7$ and 1.5 based on samples 1 and 3, and 2 and 
3 respectively. The ratio between the self diffusion constants of the two 
gases is equal to the inverse of the square root of their molar masses, i.e. 
1.47.  That the values we have found in the aerogel are so close to the 
value predicted by Graham's law, suggests that the two gases 
do not interact with the pore walls in a significantly different way. 
This was expected as we designed our aerogels in order that they should be 
as chemically inert as possible with respect to the two gases.  

We show in figure \ref{fig2} the result of a two-and-a-half hour diffusion 
experiment.  The precipitate forms a single well-defined narrow sheet.  
In order to determine whether the precipitation sheet clogs the aerogel, 
i.e.\ lowers its permeability  significantly, we used the same setup as for 
the data shown in Table \ref{tab1}.  An aerogel of length 21.6 mm was 
connected to the end of a glass tube of length 1507 mm. Thereafter a cotton 
plug with ammonia was placed next to end of the aerogel, and a cotton plug 
with hydrochloric acid was placed at the other end of the glass tube.  The 
position of the first appearance of a precipitate was 
recorded at a distance 1175 mm from the HCl end, giving a diffusion constant 
$D_{NH_3}=210$ mm$^2$/h in the aerogel.  The aerogel was then removed, sealed 
with Teflon tape and cotton plugs drenched in the chemicals were placed at 
its two open ends.  They were left in place for two hours, after which the 
Teflon tape was removed, and the gel treated in vacuum at 
150 C for several hours, thus removing any traces of the diffusing and 
reacting gases but not the precipitate.  The aerogel was then connected to 
a glass tube of length 1504 mm and a new diffusion-reaction experiment 
performed.  The NH$_4$Cl ring now formed at a distance 1182 mm from the 
HCl end, giving $D_{NH_3}=207$ mm$^2$/h.  Through the Einstein relation, the 
diffusion coefficient is proportional to the permeability. Thus, with the 
accuracy of our experimental method, a single precipitation sheet does not 
influence the overall permeability of the aerogel.  We then let a large 
number of  sheets form by continuing the diffusion process over a period of 
5.5 days.  By repeating the diffusion constant measurements for 
this system, we detected a relative decrease in the diffusion coefficient 
of approximately 30\%.  Thus, there is a detectable influence of the 
sheets on the permeability.

\subsection{The Liesegang Sheets}
\label{sect2c}

The sheet-like structures we observe in the aerogel system (figure 
\ref{fig1}) is an example of the Liesegang phenomenon.  The sheets are very 
narrow and span the
entire cross section of the aerogel rods.  The 
spacing between the sheets is typically
very narrow --- often just a fraction of a millimeter.  
However, much larger gaps 
occationally occur  in an apparently unsystematic way.  
We furthermore note that
the typical time for the occurence of the first 
sheet is a couple of hours, while it
takes of the order of a week to generate a pattern 
such as the one shown in figure 
\ref{fig1}.  We also note that the sheets appear 
to be only slighlty curved, and are
surprisingly parallel to each other.  

The sheet separation that follow from Ostwald theory 
in the usual Liesegang setup
predicts that the ratio between the position of the rings, $x_{n+1}/x_n$
approaches a constant $p$.  This is known as Jablczynski's law \cite{j23}.
Furthermore, the ratio between the width $w_n$ of sucessive rings,
$w_{n+1}/w_n$ approaches $p^\alpha$ where $\alpha=0.5-0.6$ \cite{cld94}.  
However,
we note that the {\it boundary\/} conditions that we employ 
here differ from those
studies where these laws have been found.  Thus, we do not expect these 
to hold in our case.  This we find experimentally, and, 
as we shall see in section 
\ref{sect3}, numerically.
We furthermore note that the spatial separation of the sheets 
increases with decreasing concentration
of the reactants.  We will discuss this further in the next section.  
 
As the pore size of the  aerogel 
system is of the order of 50 nm or less, i.e.\ a
tenth of the wavelength of visible light, 
a single pore filled with precipitate is 
invisible.  Thus, in order for the 
precipitate to be visible, numerous neighboring pores 
must contain precipitate.  This suggests 
that there is a effective concentration threshold 
for visibility of the precipitate.

One may then speculate whether the 
sharpness of the sheets is an optical illusion.  In
order study this, we measured the concentration of NH$_4$Cl in
the aerogel using IR spectroscopy.  
This method consists in removing approximately 
volumes of  one mm$^3$ of the areogel from various points in the 
rod, analysing these separately.
In figure \ref{fig3}, we show the result of this
study.  As is evident, not only 
is there precipitate outside the sheets,  but the
concentration of the precipitate is at 
some points surprisingly high even though no sheet
is visible there. Perhaps we are 
dealing with sheets just below the visibility threshold?

As we will see in section \ref{sect3}, the 
numerical study predicts very sharp sheets.
Thus, we may interpret the IR spctroscopy 
results as evidence of  a substructure of sheets
below the visibility threshold, rather than few and diffuse sheets.

\section{Numerical simulation}
\label{sect3}
As seen above, the structure of the Liesegang sheets found in the present 
experiment differs from those resulting from using ``traditional" 
boundary conditions.  It is the aim of this section to qualitatively
investigate the standard Ostwald theory \cite{d86} under the present boundary 
conditions.  Due to the very fast reaction between HCl and NH$_3$,    
we believe that a nucleation threshold for the reaction product, rather
than a reaction threshold, as suggested by Prager \cite{p56}, to be the
underlying mechanism.

We start by modeling the system in one dimension.  This assumption is 
justified by the fairly parallel alignment of the observed sheets, and the 
large aspect ratio of the aerogel rods. We therefore consider the following 
one-dimensional time-dependent set of equations,
\begin{eqnarray}
\partial_t a(x,t) - D_{NH_3} \partial^2_x a & = & - Rab 
\label{sys1} \\
\partial_t b(x,t) - D_{HCl} \partial^2_x b & = & - Rab
\label{sys2} \\
\partial_t c(x,t) - 
D_{NH_4Cl} \partial^2_x c & = & + Rab - N_1f(c)c^2 - N_2cs
\label{sys3} \\
\partial_t s(x,t) & = & N_1f(c)c^2 + N_2cs\;. \label{sys4} 
\end{eqnarray}
The four different functions, $a, b, c$ and NH$_4$Cl(s), 
represents the concentration of the reactants NH$_3$ (NH$_3$) and HCl (HCl), 
NH$_4$Cl in gaseous form (NH$_4$Cl(g)) and lastly NH$_4$Cl in solid form (NH$_4$Cl(s)).
While the left hand sides of the equations describe the free motion
of the species, the right hand sides describe their interactions. 
The reaction of NH$_3$ and HCl into NH$_4$Cl is assumed to be fast, i.e., 
$R \gg N_1, N_2$. The term $N_1f(c)c^2$ describes the nucleation of 
gaseous NH$_4$Cl into its solid form, while
$N_2cs$ describes aggregation of gaseous NH$_4$Cl into its solid state.
On a microscopic level, the formation of the solid state is a rather complex
process. However, as a first approximation we let
\begin{equation}
f(c) = \Theta (c-c_0),
\label{theta}
\end{equation}
where $\Theta (c-c_0)$ is the unit step function. Thus, no nucleation
takes place before the local concentration exceeds $c_0$.
The numerical calculations reveals that 
the particular choice of the nucleation term is not very important.
However, the existence of a threshold value $c_0>0$ where nucletion starts
(or which below nucleation is negligible) is very important.

We assume that that the fraction of NH$_3$ and HCl that reacts
is negligible compared to the total amount of reactants. The 
concentrations, $a_0$ and $b_0$, of NH$_3$ and HCl
outside the aerogel, are therfore assumed to be constant. 
By introducing the dimensionless variable $\xi = x/L$, 
where $L$ is the length of the aerogel, we normalize the 
diffusion coefficients to $D \rightarrow D/L^2$ and let the
aerogel be defined in the region where $\xi \in [0,1]$.
The reaction front is initially formed at $\xi_0$, where 
\begin{equation}
\frac{\xi_0}{1-\xi_0} = \sqrt{\frac{D_{NH_3}}{D_{HCl}}}.
\label{x0}
\end{equation}
The asymptotic position of 
the reaction front, $\xi_1$, is found by equality of the
incoming fluxes of NH$_3$ and HCl. 
Under the asumption that the reaction region,
$\delta \xi_f$, is small 
($\delta \xi_f \ll \xi_1,(1-\xi_1)$), $\xi_1$ is found as
\begin{equation}
\frac{\xi_1}{1-\xi_1} = \frac{D_{NH_3} a_0}{D_{HCl} b_0}.
\label{x1}
\end{equation}
For the reaction front to cover some distance,
it is assumed that the concentration outside the aerogel of HCl 
($b_0$) is higher than the NH$_3$ concentration, $a_0$.  This assumption
is based on the observation that the sheets develops in the direction of
the NH$_3$ side.  We assume  that the NH$_3$ reservoir is on the left hand 
side ($\xi < 0$) and that the HCl reservoir is on the rigth hand side 
($\xi > 1$). We set arbitrarily $b_0 = 10\times a_0$ in the following. 
Furthermore, we set the ratios between the diffusion coefficients to be
$D_{NH_3}:D_{HCl}:D_{NH_4Cl} = 1.5:1.0:0.8$ in agreement with Table 
\ref{tab1}. Note that $D_{NH_4Cl}$ has not been measured, and 
that the value chosen is based on the relative masses of the 
reactants and their product.  Furthermore, we have not taken into account that
the diffusion coefficients change as a result of the precipitate clogging
the pores of the aerogels.  We base this assumption on the small effect
found in the experiment --- see Sect.\ \ref{sect2b}.

The equations (\ref{sys1}) -- (\ref{sys4}) are solved on a 
discrete lattice with an explicit method. An explicit method
is chosen because of the fast dynamics of the reaction.
For the one-dimensional calculations the interval $[0,1]$ is divided
into 1000 cells. Typically $10^6$ time steps are necessary to follow
the process. For each time step, the calculations are done in two steps.
First, changes in concentrations in each cell due the reaction NH$_3$
+ HCl $\to$ NH$_4$Cl is calculated with a following correction
due to  NH$_4$Cl(g) $\to$ NH$_4$Cl(s). Second, diffusion takes place.

By increasing the reservoir concentrations, $a_0$ and $b_0$, by a common
scaling factor, $\gamma$, one can go from the low density regime where the 
intersheet distance, $\Delta\xi$, is much larger than the width of the sheets, 
via an intermediate regime to a dense regime where sheets start to overlap.
The decrease of intersheet distance with increasing reservoir concentrations
can be understood by equaling the production rate,
$Rab(\xi_f)\delta\xi_f$ at the reaction front, $\xi_f$, with the
flux out of the reaction zone at the nucleation threshold. This flux
is approximately $D_{NH_4Cl} c_0/\Delta\xi$, which gives
\begin{equation}
\Delta\xi \simeq \frac{D_{NH_4Cl} c_0}{Rab(\xi_f)\delta\xi_f}.
\label{DX}
\end{equation}
The width of the reaction front is $\delta\xi_f$, and we have assumed that
$\delta\xi_f \ll \Delta\xi$. 
In the derivation of equation (\ref{DX}) it is also
assumed that the motion of the reaction 
front is slow compared to the aggregation
rate at the nearest sheet. This implies 
that the profile of NH$_4$Cl(g) from $\xi_f$
to the nearest sheet is approximately linear.
One could imagine that a scaling of the reservoir concentrations,
$\{ a_0,b_0\} \rightarrow\{ \gamma a_0,\gamma b_0\} $, would give a
corresponding scaling of the production rate, 
$Rab(\xi_f)\delta\xi_f \rightarrow \gamma Rab(\xi_f)\delta\xi_f$.
However, a scaling of the reaction rate, $R \rightarrow \gamma R$,
will decrease both the values of $ab(\xi_f)$ and $\delta\xi_f$. 
Thus, one cannot expect that 
$\Delta\xi \sim \gamma^{-1}$. Figure \ref{DXfig} shows the dependence
of $\log(\Delta\xi)$ on $\log(\gamma)$ 
with the two extreme cases of $\gamma =
0.125$ and $\gamma = 4.0$ as 
illustrations. There seems to be no general power
law for the scaling of $\Delta\xi$ with 
$\gamma$, altough in the limit of low
concentrations it may look as 
$\Delta\xi \sim \gamma^{-1}$. A linear regression
on the data of figure \ref{DXfig} gives a slope of $0.75 \pm 0.03$.

It is important to notice that as long as the sheets are well
separated, the intersheet distance is not dependent on the details of the
aggregation and nucleation mechanisms. Only the nucleation threshold is
important. This implies that it is possible to find $c_0$ with knowledge 
of $a_0$, $b_0$, $R$ and the diffusion coefficients .

The dynamics of the model can be understood in the following way.
At $t=0$ NH$_3$ and HCl starts to diffuse into the aerogel. After some
time a reaction front is formed at $\xi_0$. The reaction front
then starts to move to the left, driven by the concetration
differences of NH$_3$ and HCl. After some time the concentration
of NH$_4$Cl(g) is high enough for it to nucleate into NH$_4$Cl(s). As the 
concentration of the solid increases the aggregation process
becomes rapid enough to suppress any further nuclation and the
left shoulder of the sheet is formed. As the reaction front moves away
from the sheet, the flux of NH$_4$Cl(g) away from the reaction zone drops
(the gradient in NH$_4$Cl(g) decreases). Eventually, 
the local concentration of NH$_4$Cl(g) in the reaction front exceeds 
$c_0$ and the formation of a new sheet
starts. The formation of a new sheet will stop the growth of the
first sheet. This process will then continue until the reaction front reaches
its asymptotic position, $\xi_1$. Figure \ref{timefig} shows the typical
dependence of the sheet concentrations as a function of
time. The figure shows that the birth of a new sheet stops the growth of the
previous.

From figure \ref{timefig} we can see that the time between each new
sheet is formed increases. This is closly connected with the motion of the
reaction front. A new sheet can not be 
formed before the reaction front is sufficiently
away from the last sheet (as argued above). 
For a more comprehensive understanding,
it would therefore be interesting to 
examine the time dependence of the reaction 
front position. It would not be 
unnatural assume that the density profiles are linearly
dependent on $\xi$. 
This approximation should at least be asymptotically correct when
$t \rightarrow \infty$ and 
$R \rightarrow \infty$. However, assume that the
density profiles can be approximated as
\begin{eqnarray}
a(\xi) & \simeq & a_0 \left( \frac{\xi_f -\xi}{\xi_f} \right)^\alpha \\ 
b(\xi) & \simeq & b_0 \left( \frac{\xi -\xi_f}{1-\xi_f} \right)^\alpha .
\end{eqnarray}
By balancing the incoming fluxes
with the change in the total amount of the species NH$_3$ and HCl we obtain
\begin{eqnarray}
\alpha \frac{D_{NH_3} a_0}{\xi_f}\Delta t & 
= & \frac{a_0}{\alpha+1}\Delta \xi_f + \Delta R, 
\label{rf1} \\
\alpha \frac{D_{HCl} b_0}{1-\xi_f}\Delta t & 
= & -\frac{b_0}{\alpha+1}\Delta \xi_f + \Delta R,
\label{rf2}
\end{eqnarray}
in the limit when $\Delta t,\Delta \xi_f \rightarrow 0$. The amount
of NH$_3$ and HCl that reacts and forms NH$_4$Cl(g) 
is $\Delta R$. From equations
(\ref{rf1}) and (\ref{rf2}) we find that
\begin{eqnarray}
t + t_0 & = &\frac{a_0 + b_0}{\alpha(\alpha+1)(D_{NH_3} a_0+D_{HCl} b_0)} 
\times \nonumber \\
& & \left[ (\xi_1-1)(\xi_f+\xi_1\log(\xi_f-\xi_1))+ 
\frac{\xi_f^2}{2}  \right].
\label{rf3}
\end{eqnarray}
where $t_0$ is a constant.  
Thus, the reaction front velocity scales with a factor $\alpha(\alpha+1)/2$ 
compared to the case of linear density profiles. 
The logarithm appearing in this equation predicts a slowing down of
the process which is qualitatively  
consistent with what is observed experimentally --- cf.\ two hours 
to produce the first sheet, while of the order 
of a week is necessary to generate a
pattern as shown in figure \ref{fig1}.
The solid line in figure \ref{rffig} 
shows the numerically obtained reaction
front position. The broken line represents 
equation (\ref{rf3}) with $\alpha=1.25$ 
and $t_0$ given by $t(\xi_0) - t_0 = 0$,
where $\xi_0$ is defined by equation (\ref{x0}).
It is interesting to note that the reaction front initially moves to
the right and then turns to the left. 
Such effects are discussed in Ref.\ \cite{kt96}, 
although under slightly different initial and boundary conditions.

So far the numerical model we have studied has been one dimensional.  Will
the one-dimensional structure of sheets (peaks) survive in higher dimensions
in the presence of disorder, and is it possible to understand the bending
of the sheets observed in the experiments? In order to study these questions, 
we present in the following a study of the two-dimensional version of 
Eqs.\ (\ref{sys1}) to (\ref{sys4}).  

Disorder can be caused by local concentration fluctuations of the 
reacting species and presence of inhomogeneities in structure of the aerogel. 
Also, inhomogeneities in the  initial and boundary conditions, which exist
in the experimental situation, may influence the resulting sheet structure.

We consider a system of length 1 and width 0.2 (in units of the physical 
length, $L$).  This we represent as a grid of size $500\times 100$. Each
unit cell of this grid corresponds to a square of size 
$0.1\times 0.1$ mm$^2$ in the experiment.

First, consider some fixed nucleation centers
in the aerogel, i.e., points where the nucleation threshold, $c_0$, is zero.
Such points will always influence the resulting pattern, and suppress
formation of precipitate in a distance $\Delta \xi$ from that point.
The overall influence would therefore depend on the number density, $n$,
of such centers. As a general criterion for the overall effect of such points
to be small, one should require
\begin{equation}
n (\Delta \xi)^d \ll 1,
\end{equation}
where $d$ is the spatial dimension of the system.
Figure \ref{p0fig} shows 
the resulting pattern of precipitate for some values
of the number density. The nucleation centers are randomly distributed with
uniform density.

It is also interesting to study the effect of density 
variations of the medium. In particular,
since such systematic defects may possibly describe formation of 
fairly parallel but bent sheets. Indeed, equation (\ref{DX}) gives a
clear hint that nonparallel alignment is to be expected if the diffusion
coefficient is greater on one side of the system than the other.
If density variations are formed during the fabrication 
of the aerogel, this will 
result in spatial dependence of the diffusion coefficients in all directions.
On the other hand, if inhomogeneous conditions are introduced due to 
elastic compression and
stretching in the longitudinal direction, then only $D_{xx}$ will be 
spatially dependent.
Assume that the diffusion coefficients are of the form
\begin{equation}
D(x,y) = D_0\left[ 1+\delta \sin (4\pi x/L_x) \cos(\pi y/L_y ) \right] .
\end{equation} 
Figure \ref{A0fig} (a) shows the resulting pattern due to longitudinal
compression and stretching with $\delta$ = 0.2, while figure \ref{A0fig} (b),
(c) and (d) shows the resulting pattern due to density variations when
$\delta$ = 0.2, 0.1 and 0.05 respectively. The resulting pattern due
to longitudinal stretching and compression is as expected, but when
$D_{yy}$ also becomes spatially dependent, the formed pattern deviates
significantly from the original even for small values of $\delta$.

\section{Conclusion}
\label{sect4}

We have in this paper studied periodic precipitation in a gaseous 
reaction-diffusion system where both reactants, HCL and NH$_3$, 
are diffusing into the porous matrix. This is in contrast
to the usual way Liesegang phenomena are studied.  The matrix is 
provided by a silica aerogel with approximately 92\% porosity.  
The reaction product, NH$_4$Cl, precipitates
in sheetlike structures which are surpringly narrow, 
densely spaced and parallel to each
other. There is no apparent structure as to where 
the sheets appear.  Through numerical
simulations based on the nucleation theory of 
Ostwald \cite{o97}, we reproduce qualitatively
the phenomenon. 
\bigskip

We thank E.\ H.\ Hauge, P.\ C.\ Hemmer, J.\ S.\ H{\o}ye and K.\ Sneppen
for valuable discussions. Thanks are also due to S.\ Pedersen for performing 
the IR measurements, and to E.\ Nilsen for performing some of the diffusion 
experiments. F.\ A.\ M.\ acknowledges financial support by the 
Research Council of Norway (NFR). 

\begin{table}
\caption{\label{tab1}
Diffusion constants for ammonia and hydrochloric acid in different 
aerogels.  Samples 1, 2 and 3 differ in  chemical composition from samples 
4 and 5.  However, each sample within the two groups, 1, 2 and 3, and 4 and 
5, are identical. The fourth column contains the ratio between the distances 
from the two ends of the glass tube including the length of the aerogel to 
the  point at which the precipitation ring first appeared.}

\begin{tabular}{cccccc}
\hline
Sample number&Tube length&Aerogel length&Ratio NH$_3$/HCl
&$D_{HCl}$&$D_{NH_3}$\\
&(mm)&(mm)&&(mm$^2$/h)&(mm$^2$/h)\\
\hline
1&1505&19.85&0.10 &  &135\\
2&1500&10.30&1.02 &  &118\\
3&1500&10.35&52.12&79&   \\
4&1500&20.30&0.17 &  &160\\
5&1500&13.95&0.87 &  &183\\
\hline
\end{tabular}
\end{table}
\begin{figure}
\caption{\label{fig1} 
A 2.1 cm long aerogel rod with diameter 8 mm was wrapped in teflon tape 
along the long axis. At one end of the rod a cotton plug soaked in HCl was 
placed.  At the other end, a cotton plug soaked in NH$_3$ was placed.  
After 5.5 days, the cotton plugs and teflon tape were removed.  A 
precipitate consisting of NH$_4$Cl in the form of a series of clearly 
defined narrow sheets has formed. The ammonia was placed at the 
left hand side of the rod, while the hydrochloric acid was placed at the 
other side.
}
\end{figure}

\begin{figure}
\caption{\label{fig2} 
An aerogel rod after a diffusion experiment lasting two hours and thirty 
minutes.  The narrowness of precipitation region should be noted.  The 
cotton plug containing hydrochloric acid was placed at the left end of the 
rod, while the cotton plug containing ammonia was placed at the other end.
}
\end{figure}

\begin{figure}
\caption{\label{fig3} 
Drawing of aerogel indicating where IR spectra have been recorded in order 
to determine the amount of NH$_4$Cl present there. The numbers show amount 
of precipitate relative to a maximum value.  It is clear that there is 
precipitate present also outside the sheets.
}
\end{figure}

\begin{figure}
\caption{\label{DXfig} Intersheet distance, $\Delta \xi$, as a function of 
	the scaling
	of boundary conditions, $\{ a_0,b_0\} \rightarrow
	\{ \gamma a_0,\gamma b_0\} $. The solutions with
	$\gamma = 0.125$ and $\gamma = 4.0$, corresponding to the two 
	extremes in the data, are displayed at the top.}
\end{figure}

\begin{figure}
\caption{\label{timefig}Typical time dependence of sheet concentrations.
Each curve represents the maximum concentration of NH$_4$Cl(s) in each sheet.}
\end{figure}

\begin{figure}
\caption{\label{rffig}
Time evolution of the reaction front position. The solid line represents
	 the numerical solution, while the broken line represents Eq.\ (13)
	  with $\alpha=1.25$ (see text).}
\end{figure}

\begin{figure}
\caption{\label{p0fig} Influence of randomly placed nucleation centers:
	 (a) $n(\Delta \xi)^2 = 0$, (b) $n(\Delta \xi)^2 = 0.023$,
	 (c) $n(\Delta \xi)^2 = 0.056$ and (d) $n(\Delta \xi)^2 = 0.23$.}
\end{figure}

\begin{figure}
\caption{\label{A0fig}Effects of a inhomogeneous medium:
	 (a); longitudinal compression and stretching, $\delta = 0.2$.
	(b), (c) and (d); density variations, $\delta$ = 0.2, 0.1 and 0.05
	respectively.}
\end{figure}
\end{document}